\documentclass[%
 reprint,
 superscriptaddress,
 amsmath,amssymb,
 aps,pra,
floatfix,
longbibliography,onecolumn]{revtex4-1}

\usepackage{graphicx}
\usepackage{bm}

\usepackage{xcolor}
\usepackage[colorlinks=true,citecolor=blue,linkcolor=magenta]{hyperref}

\usepackage{verbatim}

\newcommand{\nn}{\nonumber}

\newcommand{\rar}{\rightarrow}

\newcommand{\ben}{\begin{eqnarray}}
\newcommand{\een}{\end{eqnarray}}

\newcommand{\be}{\begin{equation}}
\newcommand{\ee}{\end{equation}}

\begin{document}
\title{Derivation of the Landau-Zener formula via functional equations}

\author{Chen Sun}
\email{chensun@hnu.edu.cn}
\affiliation{School of Physics and Electronics, Hunan University, Changsha 410082, China}

\begin{abstract}
The Landau-Zener formula describes the diabatic transition probability of a two-level system under linear driving. Its rigorous derivation typically relies on sophisticated mathematical tools, such as special functions, Laplace transforms, or contour integrals. In this work, we present a derivation of the Landau-Zener transition probability using a fundamentally different approach via functional equations. By leveraging integrability, we prove that this transition probability satisfies a functional equation, whose solutions establish the exponential form of the formula. The coefficient in the exponent is then determined through a lowest-order perturbation calculation. This derivation is rigorous and does not involve any sophisticated mathematics. Our work provides insights into the origin of the exponential form of the Landau-Zener transition probability, and shows that the Landau-Zener formula can be viewed as a consequence of integrability, though the two-level Landau-Zener Hamiltonian itself does not satisfy the integrability conditions.

\end{abstract}

\maketitle

\section{Introduction}

The Landau-Zener (LZ) model, or the Landau-Zener-St\"{u}ckelberg-Majorana model, is perhaps the most well-known time-dependent quantum model which is exactly solvable. Since its introduction in the four pioneering works in 1932 by Landau \cite{landau}, Zener \cite{zener}, St\"{u}ckelberg \cite{stuckelberg} and Majorana \cite{majorana}, this model has become a prototype to study nonadiabatic transitions, and it has found realizations and applications in a variety of physical systems \cite{Shevchenko-2010,Shevchenko-2023}. 
The LZ model describes a two-level system 
evolving under the Schr\"{o}dinger equation with linear driving:
\begin{align}\label{}
&i \frac{d\psi}{dt}=H_{LZ}(b,g)\psi, \quad H_{LZ}(b,g)= \left( \begin{array}{cc}
 b t   &  g    \\
g &  -bt  \\
\end{array} \right),
\label{eq:Ham-LZ}
\end{align}
where we set $\hbar=1$, $\psi=(\psi_1,\psi_2)^T$ is the state vector, $b$ is the slope of the driving (we assume $b>0$ without loss of generality), and $g$ is the coupling between the two levels. We write the Hamiltonian $H_{LZ}$ with ``$(b,g)$'' to emphasize its dependence on the parameters $b$ and $g$. Evolution of the state from $t=-\infty$ to $\infty$ is described by the evolution operator $U=\mathcal{T} e^{-i\int_{-\infty}^\infty H_{LZ}  dt}$,
where $\mathcal{T} $ is the time-ordering operator. The transition probability matrix is then obtained by taking absolute squares of the elements of $U$. Being doubly stochastic, this transition probability matrix takes the following form:
\begin{align}\label{}
P_{LZ}=\left( \begin{array}{cc}
 p  &  1-p    \\
1-p &  p  \\
\end{array} \right).
\end{align}
The transition probability to stay on each of the two diabatic levels has an exact analytic formula, namely, the LZ formula:
\begin{align}\label{LZ-formula}
p=e^{-\pi g^2/b}.
\end{align}
In the four original works \cite{landau,zener,stuckelberg,majorana}, this formula was derived by four different methods: Landau performed quasiclassical analysis in the adiabatic limit \cite{landau}, Zener employed parabolic cylinder equations \cite{zener}, St\"{u}ckelberg used the WKB approximation \cite{stuckelberg}, and Majorana employed the Laplace transform 
\cite{majorana} (see also \cite{majorana-other}). See \cite{Shevchenko-2023} for a comprehensive and thorough consideration of these four works. Other methods of derivation include contour integrations by Wittig \cite{Wittig-2005}, summation of infinite perturbation series by Rojo \cite{Rojo-2010}, and the Markov approximation by Glasbrenner and Schleich \cite{Glasbrenner-2023} (see also \cite{Glasbrenner-2024}). 
A brief summary of these methods are presented in Table \ref{table}. 
Landau's quasiclassical method has very simple steps, but it gives an indefinite constant factor before the exponent, and in this sense Landau's approach involves approximations. Another simple and elegant ``one-line'' derivation is achieved by Glasbrenner and Schleich via the Markov approximation \cite{Glasbrenner-2023}. Also note that among all these methods, only two yield not only the LZ probabilities but also the full quantum dynamics including the (Stokes) phase --- Zener's and Majorana's. This --- combined with their exactness without making any approximations (via usage of special functions and Laplace transforms, respectively) --- makes these two methods irreplaceable.

In this work, we derive the LZ formula \eqref{LZ-formula} using a fundamentally different approach --- constructing and solving a {\it functional equation}. By considering a composite system made of two LZ models and reducing it back to the two-level LZ model with the help of integrability, we derive a functional equation on $p$, solving which gives the exponential form of the formula. The coefficient in the exponent is then determined by a simple lowest-order perturbation calculation. The benefits of this approach are that it is completely exact and, at the same time, mathematically simple --- it does not use any special functions, Laplace transforms or contour integrals; only elementary mathematics on matrix calculations and simple integrations will suffice.


Below, we present this derivation in a self-contained way, which shows all the calculation steps explicitly. 

\begin{table*}[]\label{}
\caption{ Summary of different methods to derive the LZ formula. Note that mathematical complexity of each method reflects the author's subjective opinion and may vary among people. The last column shows whether the phase of quantum evolution can also be obtained by the method. \label{table}}
\smallskip
\begin{tabular}{|c|c|c|c|}
  \hline
  Method by  & Method & Mathematical complexity & Phase\\
 \hline
Landau \cite{landau}
  & Quasiclassical approach in near-adiabatic limit
  & Simple  & No
\\ \hline
Zener \cite{zener}
  & Parabolic cylinder function 
  & Medium & Yes\\ \hline

 St\"{u}ckelberg  \cite{stuckelberg}
  & WKB approximation & Complex & No \\ \hline

Majorana \cite{majorana,majorana-other}
  & Two-sided Laplace transform & Medium & Yes\\ \hline

Wittig \cite{Wittig-2005}
  & Contour integration & Medium & No\\ \hline

Rojo \cite{Rojo-2010}
  &  Summation of infinite perturbation series & Complex & No\\ \hline

 Glasbrenner and Schleich  \cite{Glasbrenner-2023,Glasbrenner-2024}
&  Markov approximation & Simple & No\\ \hline

 The current work
&  Functional equation  & Simple (but needs integrability) & No\\ \hline

  \end{tabular}
\end{table*}

\section{Derivation of the functional equation}

We first show that $p$ must depend on the two variables $b$ and $g$ solely via the combination $g^2/b$. Under a scaling of time $\tau=\sqrt b t$, the Schr\"{o}dinger equation in \eqref{eq:Ham-LZ} becomes
\begin{align}\label{}
&i \frac{d\psi}{d\tau}=\left( \begin{array}{cc}
 \tau   &  g/\sqrt{b}    \\
g/\sqrt{b}  &  -\tau  \\
\end{array} \right)\psi.
\label{}
\end{align}
In the new variable $\tau$, this is again an LZ model \eqref{eq:Ham-LZ} but with a Hamiltonian $H_{LZ}(1,g/\sqrt b)$. Since the evolution is on $t\in(-\infty,\infty)$, the range of $\tau$ is also $\tau\in(-\infty,\infty)$, so $H_{LZ}(1,g/\sqrt b)$ and the original Hamiltonian $H_{LZ}(b,g)$ must have the same probability matrix. Therefore, $p$ depends solely on $g/\sqrt b$. Besides, $p$ is independent of the sign of $g$, since one can send $\psi_2\rar -\psi_2$ which does not change the transition probabilities, but the Hamiltonian becomes $H_{LZ}(b,-g)$. So $p$ depends solely on $g^2/b$. 
We thus define $\gamma\equiv g^2/b$, and write
\begin{align}\label{}
p=p(\gamma).
\end{align}

Our goal is then to determine the function $p(\gamma)$. We begin by constructing a composite system made of two copies of the same LZ model with no interaction between them. This is a four-level system described by the Hamiltonian:
\begin{align}\label{Ham-4-state}
&H_4=H_{LZ}(b,g)\otimes \hat 1_2 + \hat 1_2\otimes H_{LZ}(b,g)\nn\\
&=\left( \begin{array}{cccc}
2 b t   &  g   & g  & 0   \\
g  &  0  & 0  & g   \\
g  &  0  & 0  & g  \\
0 &  g   & g   & -2bt\\
\end{array} \right),
\end{align}
where $\otimes$ means direct product of matrices and $\hat 1_2$ is the $2\times 2$ identity matrix. The Hamiltonian \eqref{Ham-4-state} belongs to multistate LZ models --- generalizations of the LZ model to multi-level cases whose dependencies on time are still linear. Since the two subsystems do not interact with each other, the evolution of this composite system decomposes to those of the two subsystems, and its transition probability matrix can be written as a direct product of those of the two subsystems: $P_4=P_{LZ}(b,g)\otimes P_{LZ}(b,g)$. In particular, the probability to stay on the $1$st level is
\begin{align}\label{eq:P411}
(P_{4})_{1,1}=[p(\gamma)]^2.
\end{align}

Below we focus our attention to this probability $(P_{4})_{1,1}$. We are now going to show that, besides \eqref{eq:P411}, $(P_{4})_{1,1}$ is also connected to the function $p$ in another way.

We first show that the $4$-dimensional  Hamiltonian \eqref{Ham-4-state} can be effectively reduced to a $3$-dimensional one without changing the the probability to stay on the $1$st level. For the Hamiltonian \eqref{Ham-4-state} we perform an orthogonal transformation
\begin{align}\label{Ham-4-orthogonal}
&H_{4,O} =V^{-1} H_4 V=\left( \begin{array}{cccc}
2 b t   &  \sqrt{2 g}   & 0  & 0   \\
\sqrt{2 g}  &  0  & 0  & \sqrt{2 g}    \\
0  &  0  & 0  & 0  \\
0 &  \sqrt{2 g}   &  0  & -2bt\\
\end{array} \right),
\end{align}
where $V$ presents a rotation within the space spanned by the $2$nd and $3$rd states:
\begin{align}\label{rotation}
V=\left(
\begin{array}{cccc}
 1 & 0 & 0 & 0 \\
 0 & \frac{1}{\sqrt{2}} & \frac{1}{\sqrt{2}} & 0 \\
 0 & -\frac{1}{\sqrt{2}} & \frac{1}{\sqrt{2}} & 0 \\
 0 & 0 & 0 & 1 \\
\end{array}
\right).
\end{align}
Since  the $1$st state is not involved in this rotation \eqref{rotation}, the probability to stay on the $1$st level for the transformed Hamiltonian \eqref{Ham-4-orthogonal} is still $(P_{4})_{1,1}$. Now one observe that in \eqref{Ham-4-orthogonal} all elements of the third row and column are zero, so the $3$rd state {\it decouples} from the rest three states. Recalling that we are considering only the probability to stay on the $1$st level, we can remove the third row and column of Hamiltonian \eqref{Ham-4-orthogonal} without affecting this probability, so the Hamiltonian of the original four-level model \eqref{Ham-4-state} reduces to that of a three-level model \cite{bowtie-note}:
\begin{align}\label{Ham-3-state}
&H_{3} = \left( \begin{array}{ccc}
2 b t   &  \sqrt{2 g}   & 0    \\
\sqrt{2 g}  &  0  & \sqrt{2 g}    \\
0 &  \sqrt{2 g}   & -2bt\\
\end{array} \right).
\end{align}
Denoting the probability to stay on the $1$st level for this three-level system \eqref{Ham-3-state} as $(P_{3})_{1,1}$, we have $(P_{3})_{1,1}= (P_{4})_{1,1}$, so
\begin{align}\label{}
(P_{3})_{1,1}=[p(\gamma)]^2.
\end{align}

We next show that this $(P_{3})_{1,1}$ can be further reduced to a probability of a two-level Hamiltonian with the help of integrability \cite{commute}. 
According to \cite{commute}, a Hermitian time-dependent quantum Hamiltonian $H$ is integrable if there exists another Hermitian operator $H'$ --- a ``commuting partner'' of $H$, such that the following ``integrable conditions'' are satisfied
\begin{align}\label{int-cond}
[H,H']=0,\quad \partial_\tau H=\partial_t H',
\end{align}
where $\tau$ is another parameter that $H$ and $H'$ depend on. Here $H'$ needs to be nontrivial, namely, it cannot be merely a linear combination of $H$ and the identity matrix. The integrability conditions \eqref{int-cond} guarantee that the nonabelian gauge field defined as ${\mathcal A}(t, \tau) = -i(H,H')$ has zero curvature. As a consequence, the original evolution path along time $t$ (namely, from $t=-\infty$ to $t=\infty$ at a fixed $\tau$) can be deformed in the $(t,\tau)$ plane without changing the evolution operator (as long as no singularities of $H$ or $H'$ are crossed by the deformation):
\begin{align}\label{U-LZ-deform}
\mathcal{T} e^{-i\int_{-\infty}^\infty H dt}=\mathcal{T_\mathcal{P}} e^{-i\int_{\mathcal{P} } (H dt+ H'd\tau)},
\end{align}
where $\mathcal{P}$ is a path in the $(t,\tau)$ plane with the same starting and ending points as the original evolution, and $\mathcal{T_\mathcal{P}}$ is the ordering operator defined along the path $\mathcal{P}$.


In \cite{quadratic-2021}, a commuting partner for the most general three-level LZ model (up to gauge transformations) was identified, and it was then shown using path deformation that when considering transitions from and to the $1$st level, the model can be reduced to a two-level LZ model. The three-level Hamiltonian \eqref{Ham-3-state} here is a special case of the model considered in \cite{quadratic-2021} with no couplings between the $1$st level and the $3$rd level, so only a simpler version of the analysis in \cite{quadratic-2021} is needed here. Below from the next paragraph up to Eq.~\eqref{Ham-2-state-tau-infty}, we follow closely \cite{quadratic-2021} to reduce the model \eqref{Ham-3-state} to a two-level LZ model.

This analysis begins by generalizing \eqref{Ham-3-state} to depend on $\tau$ as:
\begin{align}\label{Ham-3-state-tau}
&H_{3,\tau} = \left( \begin{array}{ccc}
2b\tau t   &   \sqrt{2\tau  }g  & 0    \\
\sqrt{ 2\tau  }g  &  0  &   \sqrt{2} g   \\
0 &  \sqrt{2} g    & -2b t\\
\end{array} \right).
\end{align}
Setting $\tau=1$ in \eqref{Ham-3-state-tau}, one would recover $H_3$ in \eqref{Ham-3-state}. Following \cite{quadratic-2021}, we can identify a commuting partner $H_3'$ for the Hamiltonian $H_{3,\tau}$ in \eqref{Ham-3-state-tau}:
\begin{align}\label{comm-part}
&H_{3}' =\left(
\begin{array}{ccc}
 \frac{g^2}{2 b (\tau +1)}+b t^2 & \frac{g t}{\sqrt{2 \tau }} & \frac{g^2}{2 b (\tau +1) \sqrt{\tau }} \\
 \frac{g t}{\sqrt{2 \tau }} & \frac{g^2}{2 b \tau } & 0 \\
 \frac{g^2}{2 b (\tau +1) \sqrt{\tau }} & 0 & \frac{g^2}{2 b \tau  (\tau +1)} \\
\end{array}
\right).
\end{align}
One can check that the integrability conditions \eqref{int-cond} are satisfied by $H_{3,\tau}$ and $H_{3}'$. Actually the precise form of $H_{3}'$ is not important; it is the very existence of a commuting partner (and therefore satisfaction of the integrability conditions) that enables the following analysis.

\begin{figure}[!htb]
  \scalebox{0.6}[0.6]{\includegraphics{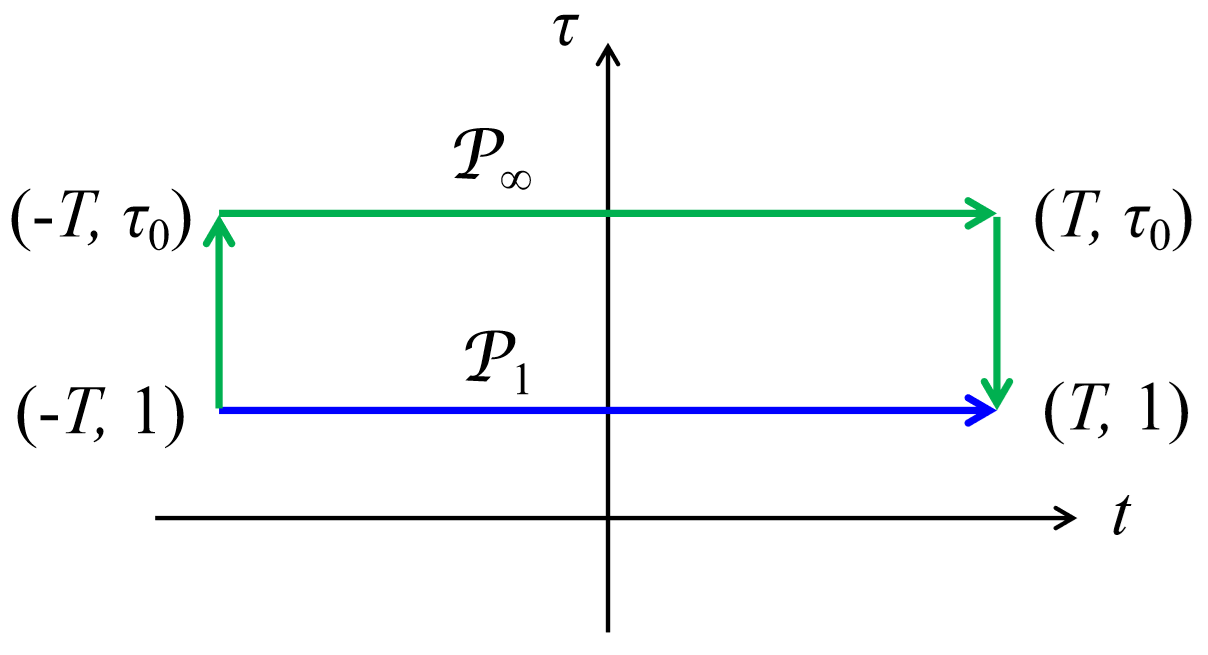}}
\caption{The original evolution path $\mathcal{P}_1$ from $t=-T$ to $t=T$ at $\tau=1$ (the blue line) for the Hamiltonian \eqref{Ham-3-state-tau} can be deformed to a new path $\mathcal{P}_\infty$ made up of three segments (the green lines) in the $(t,\tau)$ plane. Such a deformation does not change the evolution operator.}
\label{fig:path-deform}
\end{figure}

We next perform a deformation of the integration path, as illustrated in Fig.~\ref{fig:path-deform}. The original path $\mathcal{P}_1$ is a straight line from $t=-T$ to $t=T$ at $\tau=1$, where $T\rar \infty$. We deform it to a path $\mathcal{P}_\infty$ made up of three straight lines: starting from $(t,\tau)=(-T,1)$, the path first goes vertically up to $(t,\tau)=(-T,\tau_0)$, then goes horizontally to $(t,\tau)=(T,\tau_0)$, and finally goes vertically down to $(t,\tau)=(T,1)$. By integrability, the evolution operators on the two paths are identical. Now the integrals on the two vertical lines of $\mathcal{P}_\infty$ are governed by $H_3'$ at $t=\pm T$. At $T\rar\infty$, $H_3'$ is dominated by the $t^2$ term which is only in the diagonal part, so the evolutions on the two vertical lines are completely adiabatic, namely, their contributions to the total evolution operator are diagonal matrices with diagonal elements of the form $e^{i\phi}$, where $\phi$ are adiabatic phases accumulated. 
Therefore, the probability matrix associated with the original evolution $\mathcal{P}_1$ is the same as that associated with the horizontal part of the evolution $\mathcal{P}_\infty$. In other words, the probability matrices of the model \eqref{Ham-3-state-tau} at different $\tau$ are identical.

Let's now take $\tau_0\rar \infty$. Then, as argued in \cite{quadratic-2021}, the effective times of nonadiabatic transition from and to the $1$st level is $\delta t_{1,2/3}\sim \sqrt{2\tau_0}g/(2b\tau_0)=g/(\sqrt{2\tau_0}b)$, which are vanishingly smaller than the effective time between the $2$nd and $3$rd levels $\delta t_{2,3}\sim g/(\sqrt{2}b )$. Therefore, when considering the probability to stay on the $1$st level, we can neglect both the coupling between the $2$nd and $3$rd levels and the slope of the $3$rd level. In other words, to calculate $(P_{3,\tau})_{1,1}$ for the Hamiltonian $H_{3,\tau}$, one can instead use the following Hamiltonian: 
\begin{align}\label{Ham-3-state-tau-infty}
&H_{3,\infty} = \left( \begin{array}{ccc}
2b\tau t   &   \sqrt{2\tau  }g  & 0    \\
\sqrt{ 2\tau  }g  &  0  &   0  \\
0 &  0   &0\\
\end{array} \right).
\end{align}
The $3$rd state in \eqref{Ham-3-state-tau-infty} decouples, and the Hamiltonian for the other two states is
\begin{align}\label{Ham-2-state-tau-infty}
&H_{2} = \left( \begin{array}{cc}
2b\tau t   &   \sqrt{2\tau  }g\\
\sqrt{ 2\tau  }g  &  0  \\
\end{array} \right).
\end{align}
Denoting its transition probability to stay on the $1$st level as $(P_{2})_{1,1}$, we have
\begin{align}\label{P211-pgamma}
(P_{2})_{1,1}=[p(\gamma)]^2.
\end{align}
On the other band, by a gauge transformation $H_2\rar H_2- b\tau t \hat 1_2$ which does not change the transition probabilities, the two-level Hamiltonian \eqref{Ham-2-state-tau-infty} becomes just $H_{LZ}(b\tau,\sqrt{2\tau}g )$, namely, it is of the form of the original LZ Hamiltonian in \eqref{eq:Ham-LZ} with slope $b\tau$ and coupling $\sqrt{2\tau}g$. Since $(\sqrt{2\tau}g)^2/(b\tau)=2g^2/b=2\gamma$, one should have
\begin{align}\label{P211-p2gamma}
(P_{2})_{1,1}=p(2\gamma).
\end{align}
Combining \eqref{P211-pgamma} and \eqref{P211-p2gamma}, we arrive at a functional equation on $p(\gamma)$ that should hold at any $\gamma\ge0$:
\begin{align}\label{functional-eq}
p(2\gamma)=[p(\gamma)]^2.
\end{align}

\section{Solution of the functional equation}
The functional equation \eqref{functional-eq} is a special form 
of Cauchy's exponential functional equation $f(x+y)=f(x)f(y)$, since \eqref{functional-eq} is recovered by setting $x=y=\gamma$ in $f(x+y)=f(x)f(y)$. When $f(x)$ is required to be continuous, the general solutions of Cauchy's exponential functional equation are exponential functions or the zero function as discussed in \cite{Aczel}. 

Here, we will not use knowledge of solutions of the Cauchy's exponential functional equation. Instead, we assume a condition stronger than continuity -- we assume that $p(\gamma)$ is analytic at $\gamma=0$, and solve \eqref{functional-eq} by Taylor expansions in a few steps.

Analyticity ensures that $p(\gamma)$ can be expanded as a Taylor series at $\gamma=0$:
\begin{align}\label{}
p(\gamma)=\sum_{n=0}^{\infty}a_n\gamma^n.
\end{align}
Plugging this expansion into \eqref{functional-eq}, we get
\begin{align}\label{}
\sum_{n=0}^{\infty}a_n(2\gamma)^n =  \sum_{n=0}^{\infty} \sum_{l=0}^n a_l a_{n-l} \gamma^n.
\end{align}
Comparing the coefficient at each power of $\gamma$, we get a recurrence relation on the coefficients $a_n$:
\begin{align}\label{recurrence}
2^n a_n = \sum_{l=0}^n a_l a_{n-l}, \quad n=0,1,2,\ldots.
\end{align}
At $n=0$, \eqref{recurrence} gives $a_0 =   a_0^2$, whose solutions are $a_0=0$ and $a_0=1$. The case $a_0=0$ obviously leads to $a_n=0$ for any $n$, so $p(\gamma)=0$. Below we analysis the $a_0=1$ case. At $n=1$, \eqref{recurrence} gives $2 a_1 = 2 a_0 a_{1}$. With $a_0=1$, this equation is always satisfied, so $a_1$ is a free coefficient. For a larger $n$, we prove $a_n=a_1^n/n!$ by induction. Note that $a_n=a_1^n/n!$ holds at $n=1$. Suppose that $a_l=a_1^l/l!$ holds from $l=1$ to $l=n-1$ for some $n\ge 2$. Then from \eqref{recurrence}, we have
\begin{align}\label{}
&2^n a_n =2 a_n+ \sum_{l=1}^{n-1} a_l a_{n-l}
=  2 a_n+ a_1^n \sum_{l=1}^{n-1} \frac{1}{l!(n-l)!}\nn\\
&=2 a_n+(2^n-2)\frac{a_1^n}{n!} ,
\end{align}
so
\begin{align}\label{}
& a_n = \frac{a_1^n}{n!}.
\end{align}
Summing the Taylor series then gives
\begin{align}\label{}
p(\gamma)=e^{a_1\gamma}.
\end{align}
In summary, assuming that $p(\gamma)$ is analytic at $\gamma=0$, the solutions of the functional equation \eqref{functional-eq} are
\begin{align}\label{}
p(\gamma)=e^{c\gamma} \quad\textrm{or}\quad p(\gamma)=0,
\end{align}
where $c$ is an arbitrary constant. Taking into account the fact that $p(\gamma=0)=1$, the zero function solution is eliminated. Therefore, we arrive at the exponential form of the LZ formula:
\begin{align}\label{p-with-c}
p=e^{c\gamma}=e^{cg^2/b}.
\end{align}
The condition $p\le 1$ requires $c\le 0$. 
We are going to determine this $c$ by a simple perturbation calculation.

\section{Determining the coefficient $c$}

We now calculate the coefficient $c$ in the exponent of \eqref{p-with-c} by a perturbative approach. In \cite{Rojo-2010}, a perturbation expansion of $p$ up to infinite order in $g$ was obtained by skillfully evaluating high-dimensional coupled integrals using sophisticated techniques, summing which gives the LZ formula. Here, since we already established that $p$ is an exponential function, unlike in \cite{Rojo-2010}, we only need to evaluate the contributions of {\it lowest orders} in $g$ in the perturbation expansion in order to determine $c$.

Like in \cite{Rojo-2010}, we first transform the Hamiltonian into one with zero diagonal elements. Applying a unitary operator $S=\operatorname{diag}(e^{ib t^2/2} ,e^{-ib t^2/2} )$ to the Schr\"{o}dinger equation in \eqref{eq:Ham-LZ}, we have
\begin{align}\label{}
&i S\frac{d\psi}{dt}=SH_{LZ}  \psi,
\end{align}
or
\begin{align}\label{eq:Schrodinger-Spsi}
&i \frac{d(S\psi)}{dt}=\left(SH_{LZ}S^{-1} +i\frac{d S}{dt}S^{-1} \right) S \psi \equiv \overline H  S \psi,
\end{align}
where
\begin{align}\label{}
 \overline H=\left( \begin{array}{cc}
 0  & ge^{ibt^2 }    \\
ge^{-ibt^2 }  & 0  \\
\end{array} \right).
\end{align}
The new Schr\"{o}dinger equation \eqref{eq:Schrodinger-Spsi} after the transformation is on the transformed state vector $S\psi$, but since $S$ only introduces phases on the components of $\psi$, the transition probability matrix for the new Schr\"{o}dinger equation will be the same as the original one. We now expand the evolution operator $\overline U$ corresponding to $ \overline H$ by a Dyson series:
\begin{align}\label{U-overline}
&\overline U=\mathcal{T} e^{-i\int_{-\infty}^\infty \overline H  dt}\nn\\
&=\hat 1_2-i \int_{t=-\infty}^\infty  dt \overline H- \int_{-\infty}^\infty  dt_1 \overline H(t_1)   \int_{-\infty}^{t_1}  dt_2 \overline H(t_2)+\ldots\nn\\
&=\hat 1_2-i g \int_{t=-\infty}^\infty  dt
\left( \begin{array}{cc}
 0  & e^{ibt^2 }    \\
e^{-ibt^2 }  & 0  \\
\end{array} \right)+O(g^2).
\end{align}
Since
\begin{align}\label{}
 \int_{-\infty}^\infty  dt e^{ibt^2 }  =e^{i\pi/4} \sqrt{\frac{\pi}{b}},
\end{align}
we have
\begin{align}\label{}
&(\overline U)_{1,2}=-i g  e^{i\pi/4} \sqrt{\frac{\pi}{b}}+O(g^2).
\end{align}
This gives
\begin{align}\label{}
&1-p=|(\overline U)_{1,2}|^2=\left|-i g  e^{i\pi/4} \sqrt{\frac{\pi}{b}}\right|^2+O(g^3)=\frac{\pi g^2}{b}+O(g^3),
\end{align}
so
\begin{align}\label{}
&p=1- \frac{\pi g^2}{b}+O(g^3).
\end{align}
This determines $c=-\pi$ in \eqref{p-with-c}, and completes our derivation of the LZ formula.

\section{Discussions}

Having presented the whole derivation of the LZ formula, we now make some remarks on physical insights drawing from this derivation.

The first insight is the new ingredient of functional equations in treating time-dependent quantum systems. In other derivations of the LZ formula, the exponential form of the LZ transition probability usually arises from asymptotic behavior of special functions at infinite times or as results of contour integrals. Our work suggests that this exponential form can also be viewed as a consequence of solution of a functional equation (actually the initial idea that leads to the current work is that the exponential form of the LZ formula implies that it may arise naturally as a solution of Cauchy's exponential functional equation). Hence, we expect that a similar functional equation approach may also be useful when considering other time-dependent quantum models more complicated than the LZ model, given the fact that exponential forms of transition probabilities appear in many classes of exactly solvable multistate LZ models (namely, models with linear time dependence), and also in some other classes of time-dependent quantum models with other types of time dependencies. In particular, for a multistate LZ model, transition probabilities to stay on the levels with extremal slopes are described by a formula of the exponential form similar to the LZ formula, known as the Brundobler-Elser formula \cite{B-E-1993}. This formula was rigorously proved by considering analytic continuation to complex time \cite{Shytov-2004}, and by analysis of perturbation series to arbitrary order \cite{Volkov-2004}. An example of exactly solvable models with other time dependencies is
\begin{align}\label{eq:govert}
H=\left(
    \begin{array}{cc}
     \epsilon & g/t \\
      g/t & -\epsilon \\
    \end{array}
  \right),
\end{align}
which was solved in \cite{Nikolai-2014}, with a probability matrix for transitions from the eigenstates at $t\rar 0^+$ to those at $t\rar\infty$:
\begin{align}\label{}
P=\left(
    \begin{array}{cc}
     \frac{e^{2\pi g}}{1+e^{2\pi g}} &  \frac{1}{1+e^{2\pi g}} \\
      \frac{1}{1+e^{2\pi g}} & \frac{e^{2\pi g}}{1+e^{2\pi g}} \\
    \end{array}
  \right).
\end{align}
It may be interesting to pursue a new proof the Brundobler-Elser formula or new derivations of probabilities of models like \eqref{eq:govert} by a functional equation approach, though it still seems not obvious how to arrive at such derivations. 

The second insight is that the derivation shows that two-level LZ transition probabilities can be viewed as a {\it consequence of integrability}. This is somehow surprising, given the fact that the two-state LZ Hamiltonian \eqref{eq:Ham-LZ} itself actually {\it does not} satisfy the integrability conditions \eqref{int-cond} (This is because any two-dimensional Hermitian matrix do not have a nontrivial commuting partner, a consequence of a theorem of matrix commutativity \cite{Perlis-1991}.) Nevertheless, our work shows that integrability of a multistate LZ model, namely, the three-state model \eqref{Ham-3-state} helps in deriving the LZ formula. Integrability has been proved powerful in solving many integrable multistate LZ models \cite{Yuzbashyan-2018} (and solution of the two-state LZ model was used in solving those multistate models), but one might expect that the two-state LZ model itself, being non-integrable, can only be solved by methods other than integrability. The current derivation instead establishes that solution of the two-state LZ model actually also can be derived from integrability, hence enlarging further the regime of integrability down to this most basic solvable time-dependent quantum problem.

A final remark is on {\it complexity} of this approach. As illustrated above, the derivation involves only simple mathematics, mostly manipulations of matrices. This is mainly because it avoids a treatment of evolution of the state vector (or, equivalently, the evolution operator), but instead establishes a connection between the transition probabilities at different parameters ($\gamma$ and $2\gamma$) by a functional equation. As emphasized before, employment of the concept of integrability of time-dependent quantum systems plays a crucial role in reaching this connection. Now, since other rigorous derivations typically relies on sophisticated mathematical tools such as special functions, Laplace transforms, contour integrals, or summation of infinite perturbation series, can one claim that the current approach is simpler than the others? We would still prefer to answer {\it no}. This is because knowledge of integrability has to be used in this approach, which is conceptually not simple. Before the full development of this concept of integrability in \cite{commute} (this development was not at all an easy mission; before it, there were the conjecture that integrability defined as existence of nontrivial commuting partners polynomial in $t$ is necessary for solvability of a multistate LZ model \cite{Patra-2015}, and development of a set of empirical integrability conditions for solvable multistate LZ models \cite{6-state-2015}), such a derivation was impossible. In fact, a ``{\it law of conservation of complexity}'' \cite{note-complexity} is somehow effective here, and mathematical complexity (e.g. of special functions) existing in other derivation methods seem to transfer into conceptual complexity of integrability in this approach, whereas the total complexity is somehow conserved.

\section{Conclusions}

We present a derivation of the LZ formula using a method different from all previously known ones, one that involves functional equations and uses the concept of integrability.  This derivation indicates that the exponential form of the LZ transition probability can be understood as a consequence of solution of a functional equation similar to the Cauchy's exponential function equation. Our work also shows that the solution to the LZ model can be viewed as a consequence of integrability of multistate LZ models, despite the fact that the two-level LZ Hamiltonian itself is not integrable.


\section*{Acknowledgements}
The author thanks Club Nanothermodynamica and the reading club of E. T. Jaynes' book {\it Probability Theory: The Logic of Science} for organizing a systematic reading of this book; the author's initial interest in functional equations was stimulated during this reading. This work was supported by the National Natural Science Foundation of China under Grant No. 12105094, and by the Fundamental Research Funds for the Central Universities from China.


\begin{thebibliography}{}
\bibitem{note-complexity} This interesting point was brought to us by an anonymous referee; we express our thanks to him/her.
\end{thebibliography}
\end{document}